# Evaluation of Security Solutions for Android Systems

Asaf Shabtai    Dudu Mimran    Yuval Elovici
Dept. of Information Systems Engineering
Ben-Gurion University of the Negev, Israel

## ABSTRACT
With the increasing usage of smartphones a plethora of security solutions are being designed and developed. Many of the security solutions fail to cope with advanced attacks and are not aways properly designed for smartphone platforms. Therefore, there is a need for a methodology to evaluate their effectiveness. Since the Android operating system has the highest market share today, we decided to focus on it in this study in which we review some of the state-of-the-art security solutions for Android-based smartphones. In addition, we present a set of evaluation criteria aiming at evaluating security mechanisms that are specifically designed for Android-based smartphones. We believe that the proposed framework will help security solution designers develop more effective solutions and assist security experts evaluate the effectiveness of security solutions for Android-based smartphones.

## 1. INTRODUCTION
The smartphone market has grown dramatically in recent years [1]. While many players exist (for example, Google, Apple, Samsung, Microsoft, RIM, Ubuntu, and Mozilla), the major two OS players are Google (Android OS) and Apple (iOS). According to the 2013 Gartner report [1], in August 2013, the Android OS's market share was 79% followed by iOS with 14.2%. A smartphone is basically a small sized mobile computer that includes several unique subsystems such as GPS; Wi-Fi, Bluetooth, Cellular, and NFC connectivity; gyroscope and accelerometer sensors; etc. The smartphone's diverse set of capabilities opens the door to a fascinating new world of applications and new possibilities, a world that brings great comfort to the user by keeping them updated and providing accessibility to personal/business information and services, regardless of the user's location. In addition, many of the operations that were traditionally done on the PC have shifted to smartphones. The growth in the use of smartphones presents new challenges for organizations. These mobile devices are expected to be part of an organization's IT infrastructure. Because this exposes the internal network to new attack vectors, these devices should be properly integrated in a secured and safe way. This becomes more challenging in the case of *bring your own device* (BYOD), in which an organization needs to enforce its security policy on privately owned devices. The smartphone's ability to access an organization's internal network means that it may retrieve and hold valuable business data. Therefore, smartphone became an attractive target for attacks. These range from social engineering [2] attacks to hardware [3] attacks in which cyber criminals exploit vulnerabilities in human nature, software and hardware in order to accomplish their goals.

Since the release of Android-based smartphones in 2009, many security solutions for smartphones have been proposed by academia and industry. While the main focus has been on malware detection, host-based intrusion detection, access control, static analysis of applications, and encryption and isolation, recently, several solutions for system in which a secured/regulated phone that runs alongside a consumer phone on the same device have been proposed (both academic and commercial), ranging from application level solutions to hardware virtualization.

In this study, we propose an evaluation framework consisting of a set of evaluation criteria for the evaluation of security solutions for smart mobile devices. We focus specifically on the Android platform. We review several security solutions and analyze the solutions according to defined set of evaluation criteria. The evaluation criteria are viewed from the eyes of a vendor that wishes to build a secure mobile product.

The remainder of the paper is organized as follows. In Section 2 we review recent known attacks and threats on the Android OS. In Section 3 we review the set of available security mechanisms for Android. In Section 4 we present the proposed evaluation framework, and in Section 5 we apply the evaluation framework on a specific security mechanism: multiple phones running on the same device using virtualization in order to illustrate how to use our framework. We conclude the paper in Section 6 with a summary and conclusion.

## 2. REVIEW ON ANDROID THREATS
Many papers and technical reports related to Android threats have been published [4],[5],[6]. We can classify Android threats (or mobile OS threats in general) according to the methods that are used to acquire access to the device and the methods that are actually used to attack or inject the malicious code.

### 2.1 Gaining access to the device
In order to infect a smartphone with malicious code an attacker can use variety of methods to gain access to the device. These methods can be grouped into the following categories: social engineering attacks, physical access, and network attacks.

**Social Engineering Attacks**

Several social engineering attacks have been used to install malicious programs on smartphones. *Application Repackaging* is a common social engineering approach in which the attacker includes the malicious code in a legitimate application and publishes the modified application though one or more application markets [2], [7], [8]. It is usually difficult to differentiate between the legitimate application and the modified one. An attacker often uses applications that are purchased and distributed through legitimate application markets and offers free versions of the applications which include the malicious functions. During the first six months of 2011, most Android malware was spread through this method. An attacker will also package and distribute malware within a large number of applications to increase the number of infected users. Examples of such malware include the Walkinwat and DroidDream, a malware that was published with over 80 unique applications.

Another malware family that employs social engineering attacks is the *fake installer* family. Such malware appears to be a legitimate application (usually downloaded from the fake market), but actually, when executed, it installs no real application on the device, but instead installs an application acting as Trojan horse for different malicious purposes [9]. This type of attack is used in the malicious applications: FakeregsMs.a, Faketimer.a, and FakeUpdates.a.

Attackers use the fact that the protection provided by the *application permission mechanism* is not sufficient and install applications that maliciously uses permissions granted by an unaware user. This can be done by using confusing permissions definitions to hide the malicious application's true purpose, using "exotic" language when asking permissions, or using a different font size for the permissions definitions as described in [2].

Another method of spreading malware uses the *application update mechanism*. In the first step of this method an attacker releases a legitimate application. When a large number of users have installed the application, the attacker releases an update containing the malicious functionalities [2], [8]. Device owners that have set their device to automatically update their installed applications will not be aware of the update at all. This approach minimizes the exposure time of the malicious application. The first malware to exploit this technique was the DroidKungFu.

In the *drive-by download* method, the malicious application is downloaded automatically when the user browses a compromised Web page [2], [8] or when the user has their NFC turned on [10]. In some cases, the user must take action to open the downloaded application, while in other cases the installation of the application can start automatically. An example is the GGTracker Trojan which was spread though a malicious website. The GGTracker is automatically downloaded to a user's phone after visiting a malicious fake Android Market website. The Trojan is able to sign-up a victim to a number of premium SMS, subscription services without the user's consent. This can lead to unapproved charges to a victim's phone bill.

In *Phishing via SMS* the user receives a crafted SMS with a link. By clicking that link the attack is executed [11]. Sometimes this attack combines a *drive-by download* injection as well. This type of attack is used in malicious applications such as Spitmo. Spitmo's main function appears to be to intercept the single use transaction authentication number (TAN) that is communicated to the user's mobile phone during a banking transaction. It accomplishes this by intercepting all SMS messages and forwarding their content to a remote website whose URL is defined in a configuration file. The user does not get to see any of these SMS messages. Spitmo keeps a low profile by running quietly in the background and providing the user with no evidence of what it is doing. It is designed to look like it is part of the system [12]. *Malicious advertising* (or *Malvertising*) is another social engineering method in which the malware it downloaded and installed though advertisements that are presented to the user inside legitimate applications. This is a highly dangerous channel as developers use advertisements inside their applications as a business model and to gain additional users [13].

**Physical Access to the Device**

In this type of attack the attacker has direct physical access to the device. In *physical drive-by download* the attacker can use the adb debugging tool (if enabled) in order to execute various actions such as: installing malicious applications, bypassing a locked screen, making system changes, and backing up data to a remote location, and also to perform an auto token cloning attack (described later) [14], [15].

In cases in which the adb service is disabled an attacker can use the *recovery mode attack*. The recovery mode includes a separate partition on an Android device that is usually used for maintenance. When entering the recovery mode the device's regular boot sequence is circumvented. Since there is no trusted component to the boot system, attackers can utilize the recovery partition by loading their own malicious images to gain privileged access to the user's information without affecting user data. In a *recovery mode attack,* an attacker creates a customized recovery image with several modifications to files (e.g., init.rc and default.prop) in order to grant the necessary privileges. Then the attacker flushes the image to the recovery partition by running a manufacturer specific tool such as "odin" for Samsung devices [16]. After flashing the image, the attacker can access the recovery image by using a device-specific key combination and enter the recovery mode and automatically run its malicious code. fThe attacker can even gain super user rights and use the recovery mode to enter the shell and execute malicious code.

A proof-of-concept showed that an attacker can use a *modified charger* to attack the device [3]. The charger is essentially a power brick with a BeagleBoard (a low-power open-source software and hardware single-board computer) inside it. Thus, a user will connect its device to a computer rather than a simple charger. Once the charger is connected to the USB port of the device it injects the malware [17].

When an attacker has physical access to the device they can use the *cold boot attack*. In this attack the phone is put into the freezer to cool down to minus ten degrees Celsius (in which the electronic components are cold enough and the rate of data fading is slow); when the device is frozen it is taken out, and a custom recovery image is flashed to the device via "download mode." Once the modified recovery is flashed, the new operating system can start and retrieve data from the memory which contains data from the previous execution, such as the PIN number used to enter the secure area of the device [18].

**Network Attacks**

This type of malware injection exploits vulnerabilities in the mobile network architecture and protocols in order to eavesdrop and/or manipulate mobile data communications. One approach is to employ an attack via *rouge network routers* [19], [20].

Wireless networks are identified by their SSID number. When a device is registered to a wireless network (via Wi-Fi) it may ask the user to save the SSID number for automatic registration when the network will be identified again. Since the network name and its MAC address (the network interface's unique identifier) are public, an attacker can create a network with the same name (as the original network) and a fake MAC address. If the user's device is configured to connect automatically, it will register to this network when it is detected, and the attacker will have access to all data that is transmitted to and from the device.

Attacks using *rouge base stations* can occur when mutual authentication between the Subscriber Stations (SS) and the Base Stations (BS) does not occur (for example, in WIMAX, GPRS, EDGE, UMTS, or HSPA) [18]. In such cases, *rouge base stations* can generate and transmit any message to the SS.

An attacker can also use the Near Field Communication (NFC) component (if switched on) to force the Android smartphone to browse a malicious website and download malware [21], or to relay an attack on contactless transactions [22].

## 2.2 Vulnerabilities and Exploitation

The second stage, after acquiring access to the device, is the exploitation of vulnerabilities that eventually lead to the malicious code injection. In this section we review several known vulnerabilities, some of them has already been patched.

**Privilege Escalation**

There are a wide range of methods that can be used for acquiring root privileges on the device. A common approach is exploiting vulnerabilities in services/processes that run with root privileges. The *adb setuid exhaustion attack* (also known as *'rage against the cage'*) [23] is one of these methods. In this attack the attacker first identifies the processes than are allowed to run simultaneously on the system (denote NPROC) and the pid (process id) of the running adb daemon. Then, the attacker forks more than NPROC dummy processes. When reaching the NPROC, the attacker kills one of the dummy processes and then kills the adb daemon. The adb service restarts the adb daemon, and at the same time the attacker tries to reach the NPROC limit again; this leads to a race condition between the attacker and the adb daemon. When the adb daemon starts up on an Android device, it is running as root. The code will later check if it should stay as root, or run in secure mode which drops its privileges to the shell account. The attack attempts to max out the process so that when the adb daemon attempts to call "setuid()" in its code, the call will fail and the daemon will continue to run as root. The current adb code does not check if the "setuid()" call was successful or not, so it will keep running as root even if this fails [23].

In the *KillingInTheNameOf* attack the attacker exploits vulnerabilities in the customized shared memory allocator (ASHMEM) of Android in order to set adb daemon with root privileges. In the ASHMEM any user can remap shared memory belonging to INIT which contains the property space with PROT_READ|PROT_WRITE permissions and toggle properties. Basically, this type of exploit invokes the "mprotect()" system call to add write access to its mapping of the system property space, and it is then free to modify any system property at will via direct memory write. It can be used to alter the "ro.secure" bit from "1" (shell privileges) to "0" (root privileges) as described in [24].

The *udev exploit* abuses the fact that in Android a large portion of the udev code has been moved in to "init" daemon. The udev code allows standard users to "hotplug" devices that may require root level access, such as a USB device. Udev versions before 1.4.1 did not verify that these messages actually originate from the kernel. Thus, a rogue application can submit a message to udev and have an action executed (which, in the case of Android, is the init process running as root) [25].

**Third Party Library Vulnerability**

This type of vulnerability is found in the third party library and native code so an attacker will try to exploit malformed boundary checks, type safety issues, and the cross platform compile mechanism. *Browser exploits* are designed to take advantage of vulnerabilities in a browser engine or software that can be launched via a web browser (e.g., Flash player, PDF reader, or image viewer). Simply by visiting a web page, an unsuspecting user can trigger a browser exploit that can install malware or perform other actions on a device [2], [7], [26]-[28].

For example, by default the privacy setting of a camera and microphone are set as enabled for scripting. A bug in the Flash player may allow an attacker to entice users to websites that take advantage of this bug that allows an attacker to inject a script behind images that can use the camera/microphone. A proof-of-concept was shown for Linux, Windows, and Mac, and since Android supports Flash it is subjected to this kind of attack [29].

In SQL injection exploits, the attacker includes portions of SQL statements in an entry field in an attempt to get the website to execute a newly crafted SQL command that can extract data of interest to the attacker (e.g., a list of relations in the database) or modify data [30].

*Cross-site scripting* attacks may be used by attackers to bypass access controls such as the same origin policy. On mobile devices an XSS attack can inject a JavaScript XSS payload that can trigger the installation of any application on the phone when the user is browsing a legitimate website that was injected with XSS payload [28], [31], [32].

**Dalvik Vulnerabilities**

Exploiting *vulnerabilities in the Dalvik code* can lead to a denial of service of the Android instance generator. This was previously done by exploiting a vulnerability in the Zygote that enabled the attacker to force the fork mechanism and flood the device with a large number of requests for dummy instances, thus using up all of the device's memory resources [33], [34].

Another example is the *ZimperLich* exploit. Similar to the "adb setuid exhaustion attack," this attack exploits the fact that there is no check of the return value from the function "setuid()" of the Zygote process, triggered through Dalvik application components, and whose privileges are managed by Zygote. First, the exploit code forks itself repeatedly in order to reach the maximum number of processes allowed per uid (RLIMIT NPROC) for the app UID. It then issues a request to the Zygote over the local socket to spawn one of its components in a new process. When the Zygote forks a child process and attempts to set the app UID, and the setuid() call fails because the resource limit has been reached. As the Dalvik VM did not abort in this situation, execution proceeds, and the malicious application's code is running in the same UID as the Zygote, i.e., as a root process. The Zimperlich exploit then proceeds to re-mount the system partition read-write and creates a setuid-root shell in the system partition for later use in obtaining root access at any time [24],[35].

Mulliner [36] introduces the Dynamic Dalvik Instrumentation tool that can be used for hooking upon the Dalvik (via JNI and Dalvik hooking library), thus enabling it to change/modify application code (i.e., injecting code) during runtime.

The *Plankton* malware exploits a vulnerability in the Dalvik class loading capability and dynamically extend its own functionality [37]. Plankton is included in host applications as a background service. This background service is called in the modified onCreate() method of the main activity of the application. Thus, when the infected application is executed, it brings up the background service. The background service collects information such as the device ID and the list of granted permissions to the infected application and sends that information to a remote server. On the server side, the collected information is processed, and based on that information (especially the list of granted permissions), the server will return a URL for it to download. The URL points to a jar file with executable code (i.e., Dalvik bytecode). The jar file is essentially a payload, which once downloaded, will be dynamically loaded (through the standard *DexClassLoader*). Doing so will allow the payload to evade static analysis and make it hard to detect. After loading, the init() method of a hardcoded payload class is invoked (through the reflection API in Android).

Finally, a method for hiding code sections of Android executable file is presented in [38]. This is done by patching the method's declaration in the .dex file, re-computing the .dex file, and then re-building the .apk.

**Man-in-the-Middle Attack**

Such methods use a rogue application that is placed on the communication line between the user and the service consumed. The rogue application pretends to be the wanted service from the user's point of view. Thus, all information is tunneled through it, and making it accessible for the rogue application to manipulate/steal the data.

For example, malware capable of performing an MiTM attack was discovered during 2012 by McAfee Labs [39]. This malicious application targets specific well-known financial entities posing as a Token Generator application. The malware uses the logo and colors of the bank in the icon of the application, making it appear more reliable to the user. To get the fake token the user must enter the first factor of authentication (used to obtain initial access to the banking account). If this action is not performed, the application shows an error. When the user clicks "Generar" (Generate), the malware shows the fake token (which is in fact a random number) and sends the password to a specific cell phone number along with the device identifiers (IMEI and IMSI). The same information is also sent to one of the control servers along with further data such as the phone number of the device. The list of control servers is located in an XML file inside the original APK. The first two lists are used to execute the MiTM attack by filtering incoming SMS messages to get those that have mTANs. If the originating address and message body are found in the "catch" list, the content is sent to the default control server. The SMS can also be forwarded to the number specified in the XML if it is configured in the "catch" list with the attribute "toSms" [39]. Other similar malware are Zeus and SpyEye [40], [41]. An attacker can also set up a Wireless Tether to make the phone appear to be a public Wi-Fi access point to the victim, and then once it is connected to the Wi-Fi network, the attacker can capture any traffic for later analysis [42].

**Return to libc Attack**

Operating systems use several mechanisms to secure the system from exploiting stack overflow. One of these mechanisms is setting stacks to be non-executable and therefore, jumping to the shell code will cause the program to fail. This protection scheme is however, not foolproof. A variant of the buffer-overflow attack called the *return-to-libc* attack does not require an executable stack nor use shell code. Instead, it causes the vulnerable program to jump to some existing code, such as the "system()" function in the 'libc' library, which is already loaded into the memory [43].

**JIT-Spraying Attack**

Android applications run as byte code (using interpreter) and not compiled code. In order to accelerate the process, an application code segment that is currently running is preoperatively compiled to physical machine code. Thus, the JIT compiler is one of the few types of programs that can be run in a non-executable-data environment. An example of that technique involved feeding ActionScript to the Adobe Flash Player VM which contained attacker-controlled constants by XORing them together as described in [44]. In the latest Android OS, the data execution prevention (DEP) [45] and address space layout randomization (ASLR) [46] protection mechanism was introduced to prevent return oriented programming (ROP) attacks.

**Network Architecture Vulnerability**

Network exploits take advantage of software flaws in the mobile operating system or other software that operates on local (e.g., Bluetooth, Wi-Fi, NFC, DNS) [47], [48] or cellular (e.g., SMS, MMS) networks [49], [50]. Network exploits often do not require any user intervention making them especially dangerous when used to automatically propagate malware [2], [51].

**Virtualization Vulnerabilities**

Hypervisors for virtualization are becoming available for Android which makes the Android device potentially vulnerable to such attacks. A proof-of-concept from the PC platform showed that hacking the hypervisor, or gaining control of one of the virtual systems and using the "escape to hypervisor" vulnerability is possible. This vulnerability allows an attacker to "escape" from a guest virtual machine and to interfere with other virtual machines, or the hypervisor itself. For example, a virtual server's guest OS can perform lower-level system calls to the hypervisor, sometimes referred to as hypercalls. The hypervisor frequently doesn't check the hypercalls to ensure that they were invoked by the guest operating system or an application running in the virtual server. As a result, if an attack comes through a guest operating system, it may get out of the virtual server and compromise other components, including the virtual host, hypervisor, other virtual servers or other hosts [52].

**Android Debug Bridge**

When installing an application via a cable using the adb shell, no permission acknowledgement from the user is required, thus a malicious application can be granted all the permissions it requests [53]. As a case in point, on January 2014 a new malware, Trojan.Droidpak, that exploits the adb utility was discovered. This malware infects Windows machines and installs on the infected machine the adb command line tool. When an Android device is connected to the infected machine via USB it installs a malicious apk. This attack vector will work only if the "USB debugging" option on the device is enabled [54].

**Kernel vulnerability**

This exploit uses vulnerabilities within the kernel to gain user information, or root privileges [55], [56]. The kernel is major part of the Android's OS. Adding modules to the kernel increases the risk for vulnerabilities. Two known vulnerabilities within the kernel that can allow an attacker to gain user information or root privileges are discussed in [55] from the Tab's family devices' perspective. The first is that malicious programs can escalate the root-level privilege of a process, through which it can disable the security software, inject malicious codes and install rootkits. It does so by using the lack of a boundary check to set a kernel buffer overflow when calling "vhost ioctl ctrl module regrdwr" read and write sub functions (these functions are managing the kernel memory access). The second pitfall exploits the fact that there is no size check in function "BUG ON(IOC SIZE(cmd)→NVHOST IOCTL CTRL MAX ARG SIZE)". The code does not check the size of "(IOC SIZE(cmd)" which leads to system crash. Combining the two pitfalls together can generate a DoS attack. Another vulnerability was found in the Samsung Galaxy S3's kernel. The kernel allows the device to be readable and writable by all users and gives access to all physical memory thus resulting in a privilege escalation attacks [56].

# 3. SECURITY SOLUTIONS FOR ANDROID

In order to harden and protect an Android device several safeguards may be employed. Many mechanisms and solutions are already provided by security companies, and many of them were proposed and evaluated by the academic community. In the following paragraphs we provide a description of several security mechanisms that can be installed on Android-based devices in order to improve their security.

## 3.1 Anti-Malware

To identify and remove malware, anti-malware software examines files, email attachments, memory, system configuration, MMS, Bluetooth objects, etc. It usually identifies known malware based on a signature repository. Several commercial solutions are available for Android which provide an anti-malware component. There are also open-source anti-virus and rootkit detectors that can be ported to Android such as the ClamAV [57]. Anti-malware is a well-known solution and is extensively used in other platforms. At this time, the anti-malware solution does not seem to be effective for mobile devices. The reason for that is the anti-malware applications run at the same level and have the same type of privileges as any other application. Most of the currently available anti-malware applications are signature-based applications that require continuous updating of the signature repository; they provide very low false-positive rates but can detect known malware and. Rastogi *et al*. evaluated various anti-malware protection applications and their resistance to transformation attacks [58].

## 3.2 Firewalls

A firewall running on Android can prevent remote network attacks. It is a well-known and highly effective solution; however, it will not protect against attacks via web-browser, SMS/MMS, email, or Bluetooth and will not provide phone call filtering. Firewall is supported on Android through the NetFilter. The Netfilter is a Linux kernel subsystem that provides firewalling capabilities (e.g., packet filtering and connection tracking capabilities). In order to update the firewall policy, the control application should run with root privileges [5]. Recently, application-level solutions have been proposed such as the Mobiwol firewall [59] that provides a protection layer that allows applications to exert fine-grained control over the assignment of permissions through explicit policies that are similar to a network-level stateful firewall.

## 3.3 Intrusion Detection System

Intrusion detection systems (IDS) monitor the device, applications, or user's behavior to detect/prevent abnormal or known malicious behavior. Anomaly-based IDS can detect unusual phone call/SMS activity, denial of service attacks, and protect the information on the device in case of theft or loss. While it may detect new and isolated attacks, it will probably suffer from a high rate of false positives. Many academic initiatives to enhance protection of mobile devices have employed host-based intrusion detection systems comprised of an agent collecting various features from the device and the subsequent application of various algorithms to classify the behavior of the system as benign or malicious or to detect anomalies. Some solutions, such as Andromaly [60], [61] provides data collection and analysis on the device in order to detect abnormal, potentially malicious behavior on the device or an application. Other solutions, such as the Paranoid Android [62], are based on a centralized analysis that utilizes the multiple data sources and unlimited computation resources of the cloud [63]. As with any other security mechanism, the IDS should run with special privileges to ensure trusted monitoring.

## 3.4 Access Control

Android incorporates several access control mechanisms [4]. While these mechanisms are enforced on the application level or only on files, Linux can provide other tools that are directly enforced by the kernel. An example of such access control mechanism is the SELinux that was applied on Android [64], [65]. SELinux allows for the restriction of any process of the system, including root-owned, and by limiting access of processes and users to resources and/or services, the potential damage from malicious or exploited applications is limited. Its decisions are based on an access control policy, which should be deployed together with the base system. SELinux was shown to be effective against various attacks including protecting against vulnerability found in the Android volume daemon (vold) which runs as root (did not verify message origin), vulnerability in the Zygote process, Zimperlich, and even vulnerability in the Skype application [35]. Recent versions of Android provide simple authentication functionality based on a screen lock pattern mechanism. Such mechanisms can be extended so that the device can be locked remotely (when the device is lost or stolen), or by protecting sensitive information stored on the device, or on the SD card using password-based encryption. In the same context, Ni *et al*. [66] present the DiffUser framework that provides role-based access control mechanism for smartphone users. DiffUser was implemented and evaluated on Android. Each user can be assigned different rights. For example, only administrators can install/uninstall applications or the guest user can only use the phone application.

Many research papers focus on the protection of the Android permission mechanism. During the installation of the application on Android, the user may view a list of required permissions and decline installation based on this list. However, there is no way for the user to allow only a subset of the required permissions. Nauman and Khan (2011) [67] added an advanced feature to the Package Installer enabling the user to decline certain requested permissions but still permit installation of the application. Such a change would be highly beneficial to security aware users. This solution would provide protection from the possibility of granting unneeded permissions that could be maliciously used. However, applications granted with a partial set of permissions may crash if the developer did not anticipate this in advance and provide a solution for such a situation (i.e., handle cases in which partial permissions were given). This solution can be enhanced for corporate users to provide the option for hardening Android devices by limiting granting permissions based on a predefined policy. Additional efforts for enhancing Android's application level permissions mechanism are presented by the Kirin system [68] and Secure Application INTeraction (SAINT) [69]. These two systems presented an installer and security framework that place an overlay layer on top of Android's standard application permission mechanism. This layer allows applications to exert fine-grained control over the assignment of permissions through explicit policies. The XManDroid framework is used for monitoring and enforcing policy on the inter-component communication (ICC) channel [70]. The ConUCON system [71] enforces a policy that is based on contextual attributes. For example, a battery level lower than 30% will not permit games to be activated. The TISSA system implements a context-aware privacy mode that is able to control what information should be accessible to an application [72]. The TainDroid system was proposed for monitoring in real-time how third party applications access and manipulate users' personal data [73].

## 3.5 Spam-Filter

A spam filter can block unwanted MMS, SMS, emails, and calls from an unreliable origin. In the mobile phone arena spam filters are implemented using the white/black listing approach, with caller ID and a words/phrases dictionary used as the sources for allowing/blocking a call or a message. Products for spam filtering on Android are already available (e.g., SpamDrain and SMS Spam blocker). Email spam filtering can be provided by either the email server (e.g., gmail) or by an Android client side email application. Research studies also attempt to mitigate the spam problem on mobile devices. An example is the SMSAssassin [74] that applies Bayesian Network and SVM algorithms for detecting spam SMS, while keeping the model updated using crowd sourcing.

## 3.6 Automated Application Analysis

Since Android (and Google) does not provide a certification process for applications (similar to the Apple application store), many attempts have been made to apply automated application analysis (both static and dynamic). Android .apk files encapsulate valuable information that can help understand an application's behavior. This information includes requested permissions, framework methods called by the application, framework classes used by the application, user interface widgets, and more. Schmidt *et al.* [75] evaluated a framework for static function call analysis and performed a statistical analysis on function calls used by native applications. In [75], Schmidt *et al.* (2009) proposed a collaborative approach for analyzing applications and the detection of Android malware. In [76], machine learning classifiers were applied on static features extracted from Android's application files. In this approach, the application file is represented by a vector of static features extracted from the file, and the classifiers are then applied to learn patterns in the code in order to classify new files. Chaudhuri [77] presented a formal language for describing Android applications and data flow among an application's components. This formal language can be used for statically analyzing Android applications and data flow between applications and comparing those with security specifications defined in the application's manifest. This lays the groundwork for security decisions such as: is the application safe and does it do what it claimed to do. Therefore, it can provide the means for a developer to certify its application and for the user to verify the proof of the certification before installation. The Crowdroid framework [78] performs dynamic analysis of Android applications. It analyzes traces collected from many devices in an attempt to detect malicious behavior. The DroidRanger [79] uses permission-based filtering and heuristic-based behavioral footprint matching for the detection of malware in different application stores. In [80], a thorough analysis of permissions that are requested by applications is performed in order to detect potentially malicious applications. Automated application analysis is closely coupled with malware detection mechanisms and certification and can provide an automated alternative as a part of the certification process. In this way developers can certify their applications, and users can verify the validity of the certification before installation. In an attempt to reduce malware getting on the play store, Google introduce on 2012 the Bouncer service, which was shown to be easily bypassed by Oberheide and Miller [81].

## 3.7 Data Leakage Prevention (DLP)

Data leakage prevention mechanisms prevent sensitive/private content from leaking out. Identification of such content is carried out by various content and context inspection mechanisms (e.g., predefined keywords and patterns/regular expressions, fingerprinting of sensitive content and statistical algorithms). These mechanisms haven't yet been fully integrated into smart mobile platforms despite the fact that these devices can store content that should be protected (e.g., location, documents, contacts, calendar, etc.) Various research papers analyze the behavior of applications and granted permissions on mobile devices through network features and their access to sensitive data in order to detect potential leakage incidents [72]. An additional DLP feature, anti-theft, is provided for smartphones by several security vendors. This feature provides remote control capability over the device in case it gets lost or stolen. This module enables users to locate the device, block it, and wipe its data remotely.

## 3.8 Address Space Layout Randomization and Data Execution Prevention

Since the release of Android 4.0 (Ice Cream Sandwich) and Android 4.1 (Jelly Bean), Google added full implementation of two new mechanisms to enhance the system security. Address Space Layout Randomization (ASLR) is a technology used to prevent shell code from being run successfully. It does so by randomizing the base points of various areas of memory (e.g., stack, heap, libs, etc.) The goal of ASLR is to make certain classes of control-hijacking attacks more difficult. Attacks are significantly harder to develop and execute because of the following: executable code residing in unknown locations, variety buffer, or stack overflow. Data Execution Prevention (DEP) is a mechanism that prevents certain memory sectors, e.g., the stack, from being executed (also known as Nx bit). When combined, the exploitation of vulnerabilities in applications using shell code or return-oriented programming (ROP) techniques are becoming difficult in a mobile platform [46].

## 3.9 Virtualizations, Security Extensions

Virtualization is an emerging trend that is slowly becoming supported by different Android implementations, third party software stacks, and hardware implementations in order to enable multiple isolated phones (environments) to run on the same device. We distinguish between five types of virtualization: Linux virtualization (Domains), micro kernel virtualization, full virtualization with binary translation, para virtualization, and hardware-assisted (faithful) virtualization. These approaches differ primarily regarding the layers of the operating system (OS) and HW in which virtualization functionality is implemented. Full virtualization with binary translation, para virtualization and hardware-assisted (faithful) virtualization represent a historical evolution of virtualization within the server world which was not adopted by the mobile world. In this chapter we also address other topics which are not directly related to virtualization but are important components of mobile security solutions which are used in combination with virtualization solutions.

**Linux Based Virtualization**

The Linux based virtualization is pseudo-virtualization that uses the capability of *Linux Containers* (LXC) which relies on the features of *Kernel namespaces isolation* and *control groups* (cgroups) in order to create different isolated execution spaces which serve as a basis for different virtual phones. Each container can be configured to gain access to its own isolated resources such network interface (and IP address), file system and process in terms of security and resource usage [82], [83]. Solutions that fall under this category utilize a single OS kernel across all virtual phones (as opposed to running multiple OS instances) where each virtualized phone operates in its own isolated system-level container. These solutions usually implement a thin intermediate layer that manages administration of the containers, communication between containers such as notifications (an example of the architecture is shown in Fig. 1).

Practically speaking, each virtual phone runs its own Zygote process (the initial Android Dalvik process), and each application within a virtual phone is forked from the Zygote process running on the same virtual phone. The main advantage of such an approach is the ease of deployment on devices, minimal reliance on hardware capabilities, and better multi-virtual phone usability thanks to the single OS context. The disadvantage lies in the fact that the kernel which serves all virtual phones is a TCB (the whole kernel). This is similar in size to the regular kernel so if someone hacks into the kernel, all the instances are compromised at some level. Examples of vendors providing Linux virtualization include Cellrox [84] and Divide [85].

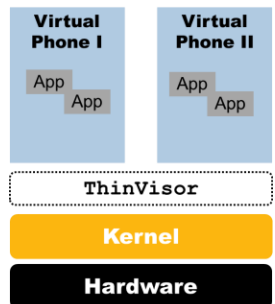

**Fig. 1**. An example of Linux virtualization implementation.

**Micro Kernel Virtualization**

In micro kernel virtualization the kernel is divided into two parts, the micro kernel and the kernel itself [86]. The micro kernel runs in privileged mode and the kernel runs in user mode. Micro kernels follow the minimal TCB principle and therefore, radically reduce the amount of code running in privileged mode by only providing the basis needed to implement functionality in user space. Critical modules that control primary resources (memory and process management and scheduling) are included in the micro kernel and are modified for virtualization. The kernel on top of the micro kernel is adapted to work with the micro kernel in order to support this separation and provides support to virtualized resources with a special driver. On top of the kernel is a specialized Android system which is adapted to the kernel runs on it. The Android is not aware of the fact that multiple stacks of kernels and Android frameworks are running simultaneously above it (see Fig. 2). Examples of vendors providing micro kernel solutions are: Trust2Core which is based on the L4 micro kernel [87] and OK Android [88].

**Hardware-Assisted Virtualization**

Hardware assisted virtualization, in which the hardware plays an important role in the virtualization tasks, is considered the most advanced form of virtualization. The hardware's roles depend on the vendor and its existing virtualization extensions where the trend includes forward improvement and enhancement of such extensions (see Fig. 3).

The hardware virtualization concept actually replaces the components implemented by para virtualization and full virtualization and thus, reduces the overhead on the software stack as well as possible vulnerabilities. Hardware virtualization has flourished in the server world and is slowly entering the smartphone world where the vendors are developing new opportunities. One example of advanced hardware based virtualization capabilities exists in the new ARM processors where you can find:

- The introduction of a new hypervisor execution mode of higher priority than supervisor mode. This will enable the VMM to execute at a higher privilege than the Guest OSs and the Guest OSs to execute with traditional OS privileges, removing the need to employ para virtualization techniques.
- The provision of mechanisms to aid interrupt handling, with native distinction of interrupt destined to secure monitor, hypervisors, currently active Guest OSs, or non-currently-active Guest OSs. This will dramatically reduce the complexity of handling interrupts using software emulation techniques and shadow structures inside the VMM.
- The provision of a System MMU to aid memory management. This will support multiple translation contexts for multiple DMA capable masters, two levels of address translation and hardware acceleration, and abstraction.
- The support in debug functionality. This is aimed at enabling debugger access to individual Guest OSs

To the best of our knowledge, as of today there is no vendor which utilizes complete hardware based virtualization capabilities in order to deliver complete hardware based virtualization for Android. This situation will probably change in the near future.

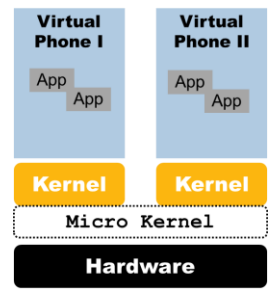 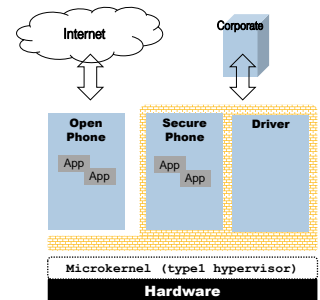

**Fig. 2**. Micro kernel virtualization.  **Fig. 3**. Hardware-Assisted Virtualization.

**Secure Boot**

The secure boot is a mechanism enabled by hardware capabilities such as the ones that exist on ARM where it verifies the integrity of boot images prior to loading them. The secure boot ensures all system software components are in a known and "trusted" state prior to running them. The secure boot process implements a chain of trust. Starting with an implicitly trusted component, every other component can be authenticated before being executed. The ownership of the chain can change at each stage - a public key belonging to the device OEM might be used to authenticate the first boot loader, but the secure world OS binary might include a secondary public key that is used to authenticate the applications that it loads.

**Containers**

A new breed of security solutions involves security containers that do not reflect a specific technique or technology, but instead are based on a similar product definition. Containers are most often regular applications (APK) that run on a device. Some containers come preloaded on the device such as in the case of Samsung KNOX [89] which only requires enabling, and some containers are fully downloadable .apks which enable the container. The functionality of containers includes (usually but not necessarily):

- Integration with a remote MDM for policy enforcement and retrieval. The container is the MDM agent on the device

which is responsible for communicating with the MDM server and verifying integrity.
- Secure filesystem in which the container replaces the file operations on the Android level (or lower level in case of hardware integrated containers) with operations that are secure.
- Possible integration with SELinux as an enforcement mechanism to prevent cross access to data and functionality which resides within the container versus outside of the container.
- Authentication and authorization of users prior to using the sensitive applications and data inside the container.
- Integration of device configuration in order to transfer MDM policy commands.

The way to implement the isolation of applications and data related to the container is container specific. Some containers have access to the ROM flashed on the device which can provide deeper integration into the OS, and other downloadable containers use available Android services or hidden ones in order to provide the suggested functionality. Two additional examples for container-based products are Mobile Spaces [90] and VMWare Mobile [91].

**Trusted Execution Environment**

A Trusted Execution Environment (TEE) is a secure area that resides in the application processor of an electronic device. The TEE ensures the secure storage and processing of sensitive data and trusted applications.

**TIMA**

TIMA is Samsung proprietary solution which is based on TrustZone and is responsible on continuous verification of the integrity of the kernel. The verification process runs inside the TEE, together with the secure boot, and ensures a strong form of protection from software based vulnerabilities which result in modifications to the kernel's storage or memory.

Solutions that fall under the trusted computing environment are the ARM Cotex-A9 [92], Quallcom Snapdragon [93], TRUSTONIC [94], and Cryptocell [95].

## 3.10 Mobile Device Management (MDM)

Mobile devices management is not a specific security technology but rather a category of products that provide tools to organizations which assist them in deploying configuration policies and monitoring these policies, and provide other centralized tasks.

## 4. EVALUATION CRITERIA

In this section we describe a set of criteria for evaluating security solutions and components for Android-based mobile devices. This set of criteria is used later on for evaluating several security components. Creating the criteria list is based, first and foremost, on the initial perception of the evaluator; in our analysis we have taken the role of a company/organization interested in building a secure mobile solution as well as using it.

**Visibility**

The *visibility* criterion indicates the amount of effort that should be invested by the attacker in order to identify the existence of the solution on the device and further extract information about it. Identification can be achieved by all means starting from running specific software on the device, inspecting the device physically, monitoring network traffic, etc. The visibility criterion can be divided into the following aspects:

- **Physical Visibility** – Indicates whether the security solution can be observed by an attacker holding the secured device, either by looking at the device or by physical (visual) inspection of the device; for example, the solution may use specific additional hardware (e.g., special micro SD card, or battery) that is noticeable physically.
- **Runtime Visibility** – Indicates what level of knowledge and inspection is required in order to notice and learn about the security solution via a piece of code running on the same device. Here we have a split analysis in which we evaluate the visibility from an application which owns root privileges or does not. For example, is viewing the task manager sufficient enough to detect the security solution? This criterion includes also the ability to detect the security solution via the available device features (e.g., camera, Wi-Fi modem) or unique setting to device peripherals, as well as deviation from normal behavior such as fast battery drain or high CPU usage at idle time.
- **Network-based Visibility**– Whether the security solution can be detected through analysis of the network traffic by finding unique patterns of the solution, such as extensive network activity with regard to the application running at the moment or unusual network settings in the device (e.g., open ports). This criterion is also divided into visibility from within the device and visibility on network elements where traffic originating to and from the device passes through.
- **Replicating Detection** – The detection effort of a previously detected solution criterion represents the effort needed for identifying another device installed with the same security solution after a single device has been compromised.

**Security Solution Administration** – Reflects various aspects of provisioning, removing, managing and controlling the security solution and/or the device.
- **Remote Control** – Is this possible and how to manage and control the security solution remotely through a network connection**.**
- **Switching Time** – The time it takes to move from the non-secure area to the secure area assuming the security solution provides such areas. For example, can it be done with a simple click or, does it require a restart of the device, and how long the switch takes.
- **Installation and Updates** – What is the method of installation, how long it takes to install, can it be installed from a remote location, how does the security solution receive updates.

**Inherent Cost**

The cost criterion evaluates the level and type of resources that are required for implementing the solution where the main driver for cost is what kind of code modifications are required (kernel, Android framework, device drivers) in order to implement the solution and which third party components are available for reuse. The second aspect of the cost criterion is migration cost where we will evaluate the cost to migrate the solution to devices from other vendors as well as new devices from the same vendor.

**Security Level**

The last and probably most important criterion is the security criterion. It reflects the level of security the solution provides,

what types of threats are mitigated, how easy it is to bypass/attack the solution; what is exposed if such a device is captured (only the solution or also the data on it). Analyzing the security level per solution is no single dimensional since it depends significantly on the attack target and vector. We will present examples of attack targets and vectors and inspect them in light of a specific solution. This can serve as a template for anyone interested in evaluating a specific solution for its own needs.

**Miscellaneous**

The **data flow criterion** refers to any supported functionalities for handling data stored on the device and how data can be transferred from one environment to the other.

# 5. SOLUTION EVALUATION

In this section we demonstrate the application of the proposed evaluation framework (i.e., criteria). We chose Linux based virtualization and micro kernel based virtualization technologies as evaluation candidates.

## 5.1 Evaluating Linux-based Virtualization

### Visibility

**Physical Visibility**

As a software-only implementation there is no visual/physical indication which can be observed on the device. Some commercial products which rely on this type of virtualization and are targeted at work/life use do provide on-screen visual indicators. There is also the risk in specific Linux containers a specific set of hardware resources will not be configured as accessible, and as such, Android itself will present specific indications (i.e., no network) which can raise suspicions; however, but such indications do not imply in 100% the existence of the solution which can be caused by other technical reasons.

**Runtime Visibility**

<u>Rooted phone</u>: In a rooted phone these mechanisms are visible based on changed device drivers, enablement of the capabilities required for Linux Containers on the kernel. The root enables immediate access to the environment of a specific container though there are attacks which enable identification of the containerization itself. SELinux enabled on a Linux based Virtualization can remove the risk of visibility in rooted mode.

<u>Non rooted phone</u>: Each virtual phone can be configured with different access to device features (GPS, Camera, network, etc.) and special configurations which are not customary on specific phone models and OS versions and this discrepancy from a standard model configuration can raise alertness as for to the existence of such solution – i.e., hiding the GPS hardware can lead to a "strange" configuration from a running software point of view. Rooted and non-rooted phones can raise this suspicion level.

<u>Rooted and non-rooted phones</u>: The CPU and battery overhead of such a solution is low and barely noticeable from a UX performance point of view. The battery consumption is mildly affected by the Linux virtualization solution since unlike other solutions (which have more complex virtualization tasks) here we have a single kernel running with optimized scheduling of resources among processes. The CPU overhead is also minimal thanks to the simplicity of multiplexing and today's common multi core architectures.

Linux containers also provide the option to "freeze" specific containers, so in a specific virtual phone setup where the foreground container is the only one that is active, the impact on the user will be almost negligible. When the background containers are active, running tasks which consume a high number of CPU cycles, it may raise the suspicion level.

**Network-based Visibility**

Each virtual phone has its own MAC address, based on the network namespace. The virtual phone does not force any special networking behavior, and as such, it can't be detected by unaware network analysis. A risk point exists if someone is able to observe the data stream of the two different virtual phones while being able to verify that both streams arrive from the same physical device (a single device in the network). For example, if the user identifies concurrent push connections to Google servers where there is supposed to be only one (one per Android instance) then this can be a suspicious indication of virtualization. In general, from its own perspective, a device with a VP solution behaves normally, meaning the network behavior is exactly the same as in a regular ROM with multiple processes using the network.

From the inner perspective of a software (non-root) monitoring the network inside a specific container the existence of virtualization cannot be deduced. A piece of software which has root privileges can monitor networking at a lower level of the networking stack where the mere existence of the network multiplexing on a single hardware device can be identified.

### Security Solution Administration

**Remote Control**

Existing Linux virtualization implementations do not support remote management. Commercial products have implemented the capability to manage containers and created integration into MDM solutions in order to manage groups of devices. From a technical perspective, controlling other containers is possible either via a container with high privileges or via the root container which controls all other containers.

**Switching Time**

Switching between containers can be accomplished in many ways (through an application, Android trigger, etc.) This is very fast and primarily depends on additional administrative processes built into the process of switching instances. The switching time is very close to the time it takes to achieve a process switch on the processor and as such, it is almost imperceptible. This estimation is based on the assumption that the containers are active and not frozen in the background and that there are no special actions triggered by the switch, such as encryption and decryption of memory/storage.

**Installation and Updates**

A basic implementation of Linux based virtualization requires building the Linux kernel after setting it in a special kernel configuration. It also requires modifications to the Android framework and modifications to some device drivers which need to be aware of such virtualization. This set of changes requires a custom ROM type of deployment scheme where installation is done via physical access. Over the air (OTA) updates are possible for such solutions although they may open a new attack vector due to the fact they enable OTA to critical parts of the operating system which can be abused by an attacker for malicious purposes. The alternative to OTA updates is, of course, secondary custom ROM upgrades on the device.

**Inherent Implementation Cost**

Creating a basic Android virtualization based on Linux Containers requires a vast amount of resources. The implementation requires modifications at all levels of the stack and only within the kernel is a reuse of the virtualization capability possible. Any advanced implementation of Android virtualization will also require modifications to the kernel base capability.

The migration effort to adapt the solution to new devices within the same product family or to support a whole new device category depends on the changes between the different platforms in these areas: the kernel, device drivers, and Android framework. These changes along with new versions or new devices dictate the level of changes required in the solution. Usually, within the same product the changes are minimal except for regular Android versioning and changes in device drivers due to incremental changes in the hardware specifications. Still, the leap between product generations can be so large that it will reflect an effort similar to new implementation of a new device family.

**Security**

**Top-down Analysis**

Analysis of the security level of such a solution is a complicated process and depends significantly on specific implementation. Commercial vendors using such technology usually fortify it with additional security extensions which dramatically change the picture in terms of vulnerabilities. We will assume for this analysis a plain vanilla implementation, which takes out-of-the-box Linux based virtualization and implements only the needed modifications on the Android level and device drivers to maintain basic operation.

We will take a top down approach in which we will first examine the attack from a non-rooted application running on an Android container.

Android no-root: The challenges for an attacker on this level are the same as the challenges of an attacker running on any Android. The vulnerabilities that exist on the same version and/or in combination with the specific hardware will also be available here. Additional risk is created based on the unique changes implemented at the Android level and device drivers in order to enable the virtualization. Such changes can expose new vulnerabilities, as well as disclose existing vulnerabilities by mistake or on purpose. Once a vulnerability is found and is being exploited, the impact depends on the exploit target. Here we assume that the exploit target was privilege escalation, and the running process/code received root permissions.

Process with root: Up until version 1.0, a process running in a Container whether secure or not can gain root privileges using common Linux vulnerabilities.. If the attacker is running in the secure area then the root privileges are sufficient to access all the sensitive material so no further attacks are needed depend on the setup and configuration of the container. In cases in which there is true isolation between the resources allocated to the secure and non-secure containers, an exploit towards the area of container administration will be required as it is the only way to access control of the secure container. Of course, the root enables other attacks which can trace networking, for example, or activating side channel attacks to deduce related information or information that belongs to the secure area. Still, such attacks will have lower impact than an attack which eventually gains control of the secure container and its assets. In general, since the Linux Containers technology is based on a shared kernel, the options for exploiting the system are wide. Following the 1.0 release on Feb 2014 the LXC concept has gained an important improvement - the user level Containers. By implementing this concept, gaining root privileges inside a container is impossible and even escaping the container does not allow immediately root privileges.

**Data Flow**

Secured data flow can be provided from one container to the other. This can be implemented in various ways including a dedicated secured inter-process communication channel. Communication between different containers via shared storage resources is, of course, possible, as well as via regular connectivity facilities of a specific container with a host computer (sync etc.). Any communication or sharing capability between containers increases the risk of finding vulnerabilities in such mechanisms and poses a higher risk on the overall mechanism where such vulnerabilities can be comparable to receiving control of the containers due to code injection techniques. An attack can abuse the sharing/communication mechanism to inject code to the secure area and gain control.

## 5.2 Micro Kernel Virtualization

**Visibility**

**Physical Visibility**

As a software-based solution the micro kernel does not have physical identifiers and cannot be detected by visual methods alone. This excludes any UI changes in the boot process as well as in the Android stack, which can be implemented as part of such a solution. In the case of the UX which provides a visual means of switching among the instances, such details can point to a specific solution. In specific implementations the recovery mode and behavior can be modified to eliminate options for overriding the implementation, and as such, it can be a visual signal of the existence of the solution. Also, special hardware configuration within specific instances can cause Android itself to either behave differently or provide different visual indications although such changes do not imply directly to a specific solution, and they can be derived from other circumstances.

**Runtime Visibility**

Special hardware configuration that is not customary to the specific hardware mode can be a suspicious sign of virtualization on rooted and non-rooted devices. The virtualization overhead in this case can be detected (with false alarms) using timing attacks at runtime in a non-rooted mode; while in rooted mode, several virtualization artifacts can be used for detecting the virtualization, such as special configuration keys, storage names, network card addresses, etc.

The battery consumption is affected by micro kernel solutions and the waste can go from 30% to 60% depending on the implementation. The impact can be attributed due to the fact that the software manages the virtualization by itself, as well as the management of IO switching between instances. Complexity in the implementation of power management for a virtualized environment can be the reason for such overhead. The CPU usage should not be dramatically high except in situations in which many context switches occur, for example, in IO intensive activities which span multiple instances. Today, with multi core CPUs the overhead of context switching of micro kernels is minimal.

The micro kernel architecture does not have an impact on networking in terms of traffic overhead. These assumptions can impact the suspicion level which can be raised by a code running and monitoring these parameters. Still, it will always raise suspicions but never directly imply the existence of the underlying mechanisms.

**Network-based Visibility**

The micro kernel does not force any special networking behavior and as such it cannot be detected upon network analysis. The micro kernel does enable, in theory, the behavior of multiple instances using the networking resource at the same time. However, in practice, the network has to be locked before each instance uses it, so eventually it behaves like two applications that are using the network at the same time. It might be the case that each virtual phone implemented over the micro kernel exposes its own unique information (such as phone numbers on one phone that are not used on the other). Still, identifying such discrepancies requires that monitoring software be installed on multiple instances in order to conduct such a comparison. An application which has root privileges cannot intercept communication arriving from different instances due to the fact the networking stack is not shared. Unless a vulnerability is found at the hardware level or device drivers which can serve as a breach to the infrastructure of the micro kernel and if such vulnerability is found then a multiplexing behavior can be identified on such level.

## Security Solution Administration

**Remote Control**
In a micro kernel setup a specific OS instance can be set with privileges high enough to control resources available to other instances, and this, combined with a proprietary control mechanism implemented in all instances (remote control agent), can be the basis from which a solution of remote control can be devised. Unlike the Linux Containers solution there is no base capability to start and stop containers or to administrate them; the setup of the instances is wired into the custom ROM containing the device image. Building a remote control mechanism which will control the instances on a micro kernel based phone defeats the isolation achieved by the mechanism and places all defenses at high risk. This is due to the fact that an attacker can exploit such mechanisms from the outside or the inside and gain control of important assets.

**Switching Time**
The switching time between instances is unnoticeable, because all of the instances are active and available for scheduling via the micro kernel. In the transition between instances there isn't a process of snapshotting and restoring the instance state. This is because each instance is in a state of activity in which its resources are being mapped differently, and each instance is associated with different permissions. The switch operation does only remapping of the resources into the new instance and as such do not take long time.

**Installation and Updates**
Since a micro kernel based environment involves changes on the kernel, the level of the drivers, and the Android framework, a custom ROM is required, and its deployment and update process is similar to any ROM update and installation procedure. An OTA mechanism for the software which resides within a specific instance is possible for implementation although its capabilities will be only replacement of applications and services and not changes in the OS stack. Core functionality, such as the kernel, device drivers, or the Android stack would not be included within these boundaries. From a software composition perspective each instance can have its own unique software stack within each OS instance in which, for example, one instance can run Linux alone, another instance can run Android 4.3, and a third can run Android KitKat (4.4), as long as all of the software stacks are adapted to run within the micro kernel framework.

**Inherent Implementation Cost**
Implementation of a micro kernel based solution renders a high cost since it requires implementation of a micro kernel stack (reuse from an open source micro kernel is possible although many additions are required to reach an operational code base), implementation of device drivers and device driver wrappers, and modifications to Android itself. Deep involvement into the device drivers makes the solution highly dependent on specific chipsets, and hardware vendors in general, which makes migration to other vendors of high cost as well. Each manufacturer can use different chipsets and/or specialized hardware, which require different drivers. Another parameter which impacts the complexity of device driver adaptation is the market position of the micro kernel vendor in which access to proprietary information related to the drivers can have dramatic impact on implementation and product testing costs. There is a big difference between adapting the micro kernel behavior towards a known device driver with a good understanding of its underlying function versus a device in which the code is unknown and the virtualization is required to generalize itself in order to adapt to the general behavior of the device.

With regard to other Android versions, the changes required for implementation depend primarily on the changes to the kernel and changes in the functionality of Android itself.

A high cost entailed in the process of building such a product is product testing which eventually includes a drastic modification to the basic operation of the device, and given this, it has to be tested as if it is a new OS release. Of course, the scope of testing depends on the target usage patterns and the size of the target audience.

## Security

**Top-Down Analysis**

As in the Linux virtualization analysis we will assume a plain vanilla implementation, which assumes an out-of-the-box micro-kernel based solution (there is no such thing as an out-of-the-box micro kernel, since all implementations are specific, but we will imagine an implementation which is based only on the capabilities provided by the micro kernel concept itself).

We will take a top down approach in which we will first examine the attack from a non-rooted application running on an Android instance with a basic setup of a secure OS instance which runs in parallel to a non-secure OS instance. Our application runs in the non-secure instance.

<u>Android no-root:</u>

In the scenario of an Android running within a micro kernel OS instance the vulnerabilities possible are the same as in any other Android, except for two special cases: new vulnerabilities inside the changed device drivers/kernel as a result of the micro kernel implementation and new vulnerabilities which can exist due to the special hardware setup assigned to this specific OS. For example, in the latter case, if the OS instance does not receive access to the network driver and network activities, this special configuration can lead to consequences and suspicious behavior in the kernel (special crashes) which can open up a new set of vulnerabilities. Once a vulnerability is exploited, the range of actions which can be taken by the exploit is dramatically smaller and confined to the assets and resources associated with the current OS instance. Even if there is an exploit which provides escalation of privileges and access to the kernel, the micro kernel lower layer protects the resources. Unless it is hacked, other OS instances and their respective resources stay untouched. Having said that, specific commercial solutions may implement different mechanisms which can create other access points to the secure OS instance. These can become weak links in the chain which can make the life of an attacker easier in that the attacker can make use of these access points instead of trying to hack and attack the micro kernel itself.

Reaching a root inside the non-secure instance does not present an immediate risk to the secure instance, and one needs to attack the micro kernel itself in order to reach other OS instances. Micro kernel related attacks are not available publicly yet, and currently, the primary preference is for denial of service attacks on the micro kernel [96].

**Data Flow**

Different instances running on a micro kernel can communicate with each other as if they are two different operating systems running on different devices. The micro kernel itself does not provide a special means of communicating among instances, although commercial products include such an option to enable different applications. This type of a channel, as well as any other option to communicate among instances, creates a risk of penetrating the secure area using protocol vulnerabilities and option code injection.

Table. 2. Comparing Linux virtualization vs. microkernel virtualization.

| Criterion | Linux Virtualization | Micro-kernel Virtualization |
|---|---|---|
| Physical Visibility | None | None |
| Runtime Visibility **Without root** | Highly Limited | Highly Limited |
| Runtime Visibility **With root** | Easy | Difficult |
| Network based Visibility | Highly Limited | Highly Limited |
| Remote control | Possible | Possible |
| Switching time | Unnoticeable | Unnoticeable |
| Installation and updates | Custom ROM + Partial OTA Possible | Custom ROM + Partial OTA Possible |
| Inherent Implementation Cost | High Cost | High Cost |
| Security level | Medium | High |
| Data Flow | Custom Mechanisms | Custom Mechanisms |

## 6. DISCUSSION

In this study we propose a set of criteria for evaluating security solutions for Android-based mobile devices. Evaluating a security solution depends significantly on the needs of the evaluator and the role the security solution will play. In this document we take the role of a company which wants to build/own a secure Android solution and wants to understand the impact of the decision to buy versus build based on criterions such as cost versus control on security. We demonstrate the evaluation framework on two new types of phone virtualization that represent different directions for virtualization that exist in commercial products today and serve as the basis for groups of products: Linux based virtualization and micro kernel based virtualization. Table 2 presents a summary of the comparison between the two virtualization methods. Another product group which provides virtualization like experiences include products which are categorized as general containers but because they don't share a similar design principle or implementation direction, evaluating them as one coherent group will not reflect all of the products' specific details. This evaluation framework can be extended and parts of its criteria substituted with alternative criteria in order to be adapted to different evaluation targets. For example, the Inherent Implementation Cost criterion is not relevant when an enterprise is looking to adopt a mobile security solution since the enterprise just adopts a security solution and does not implement one. For an enterprise the aspect of group based policy enforcement is of higher interest in order to streamline security processes.